\documentclass[%
 reprint,
showpacs,preprintnumbers,
 amsmath,amssymb,
 aps,
 prb,
]{revtex4-1}
\usepackage{graphicx}

\begin{document}


\title{Quantum Breathing Mode of Interacting Particles in a \\One-dimensional Harmonic Trap}



\author{Jan Willem Abraham}
\email[]{abraham@theo-physik.uni-kiel.de}
\author{Karsten Balzer}
\author{David Hochstuhl}
\author{Michael Bonitz}
\affiliation{Institut f\"ur Theoretische Physik und Astrophysik, Christian-Albrechts-Universit\"at zu Kiel, Leibnizstra\ss{}e 15, D-24098 Kiel, Germany
}


\date{\today}

\begin{abstract}
Extending our previous work, we explore
the breathing mode---the [uniform] radial expansion and 
contraction of a spatially confined system. We study the breathing mode across the transition from 
the ideal quantum to the classical regime and confirm that
it is not independent of the 
pair interaction strength (coupling parameter). 
We present the results of time-dependent Hartree-Fock simulations
for 2 to 20 fermions with Coulomb interaction and show how the
quantum breathing mode depends on the particle number. We 
validate the accuracy of our results, comparing them to
exact Configuration Interaction results for up to 8 particles.
\end{abstract}


\maketitle


\section{Introduction}
Harmonically confined few-particle quantum systems and especially
their time-dependent properties
are an important subject of experimental
and theoretical research activities.
For example, correlated electrons in metal clusters \cite{baletto} or quantum
dots \cite{filinov1, filinov2, reimann} 
and ultracold Bose and Fermi gases in traps or optical
lattices \cite{bloch1, giorgini} have been investigated
in recent years.
Particularly, Bose-Einstein condensation in low dimensions \cite{goerlitz} and
nonideality (interaction) 
effects \cite{menotti, moritz, pedri, bloch2}, including
superfluidity and crystallization \cite{filinov3},
lately raised attention.

Among these properties the behaviour of the breathing mode (BM) attracts special
interest, as it is easily excited experimentally \cite{moritz} and turns out 
\cite{bauch} to give information on a variety of system properties.
The BM describes a radial expansion and contraction of a finite system and 
is characterized by two frequencies in the general quantum case. 
In our previous work  \cite{bauch} we have shown for a 2-particle
system that one of those frequencies changes with the
system dimensionality, the particle spin and the strength of the pair interaction.
These results were extended \cite{hochstuhl} to 4 and 6
particles and to different inverse power law
interaction potentials \( w(r) \propto r^{-d} \)  with \(d=1,2,3 \).
In this paper we present new results for larger particle numbers
and we show how for different coupling strenghts the corresponding breathing
frequencies depend on the particle number.
For that purpose we present the results of time-dependent
Hartree-Fock simulations for up to 20 particles and compare them to
the results of exact CI results for up to 8 particles.

\section{Theory}

\subsection{Time-dependent Schr\"odinger equation}
\begin{figure}[htb]
\begin{center}
  \includegraphics[scale=1]{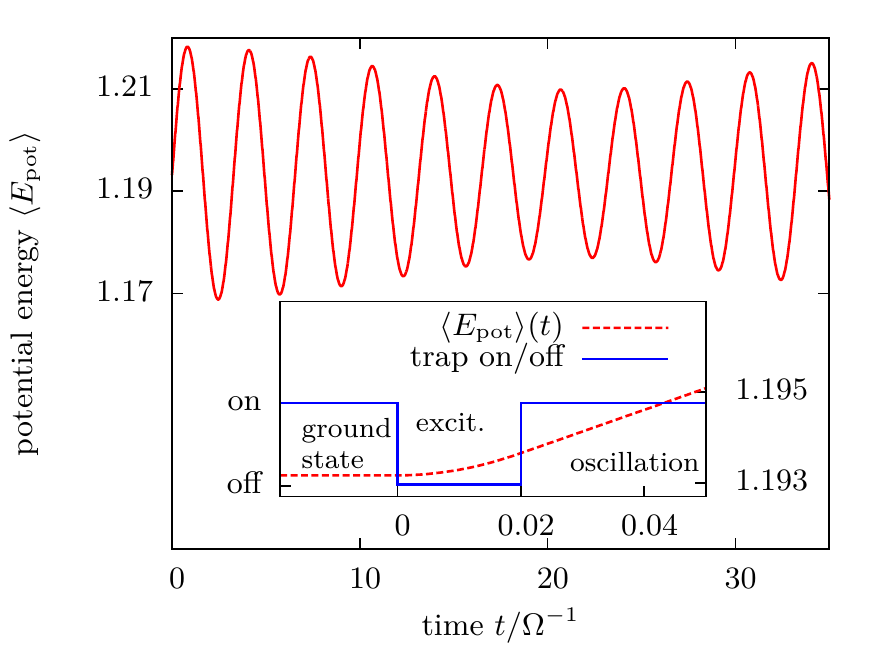} 
  \end{center}
    \caption{Exemplary time series (\(N=2,\; \lambda=1 \)) of the potential energy \( \langle E_\mathrm{pot} \rangle \). The waveform shows a superposition of two harmonic oscillations with frequencies \( \omega_r \) and \( \omega_R \). The inset demonstrates the excitation of the breathing mode.}
    \label{fig:zeitserie}
\end{figure}

We briefly recall the theoretical background of the BM \cite{bauch, hochstuhl}.
Generally, a system of \( N \) interacting particles can be described by the Hamiltonian 
\begin{equation}
	\label{eq:tidhamiltonian}
 \hat H_0 = \sum _{i=1}^N \hat h_i + \sum_{i < j}^N w(|\mathbf{r}_i - \mathbf{r}_j|) \;,
\end{equation}
where
\begin{equation}
 \hat h_i = \hat t_i + v(\mathbf r_i)
\end{equation}
is the single-particle Hamiltonian and \( w(|\mathbf{r}_i - \mathbf{r}_j|) \) is a
binary interaction potential. In this case,
the external single-particle potential \( v(\mathbf r_i) \) is
chosen to be harmonic, i.e.
\begin{equation}
 v(\mathbf r _i) = \frac{1}{2} m \Omega^2 \mathbf r_i^2 \;.
\end{equation}
\(v\) serves as a trapping potential with the trapping
frequency \( \Omega \). In the following, we concentrate on
Coulomb-interacting particles with equal masses \(m\) and
equal charges \(e\), e.g. electrons.
Thus, the interaction potential has the form
\begin{equation}
 w(|\mathbf{r}_i - \mathbf{r}_j|) = \frac{a}{|\mathbf{r}_i - \mathbf{r}_j|} 
\end{equation}
with \( a \equiv e^2 / (4\pi \varepsilon_0 )\). Finally, using the notation \( \mathbf r \equiv (\mathbf r_1, \ldots, \mathbf r_N ) \), the \(N\)-particle 
time-dependent Schr\"odinger equation (TDSE) reads 
\begin{align}
\mathrm i \hbar \frac{\partial}{\partial t} \mathrm{\Psi}(\mathbf r, t) =& \Bigg[\sum_{i=1}^N
\bigg(\frac{-\hbar^2}{2m}  \frac{\partial^2}{\partial \mathbf r_i^2}
+ \frac{1}{2} m \Omega^2  \mathbf r_i^2 \bigg) \nonumber \\ 
&  
+\sum_{i < j}^N \frac{a}{|\mathbf{r}_i - \mathbf{r}_j|}   \Bigg] \mathrm{\Psi}(\mathbf r, t)  \;.
\end{align}
For convenience, we introduce the scaled 
quantities \( \tilde{\mathbf r}_i = \mathbf r_i / l_0 \)
 and \( \tilde t = \Omega t \), so that after omitting the 
 tilde symbol, 
 the TDSE can be written in the  dimensionless form
\begin{align}
\label{eq:schroedrescaled}
 \mathrm i \frac{\partial}{\partial t} \mathrm \Psi(\mathbf r, t) =& \Bigg[ \frac{1}{2}
  \sum_{i=1}^N \bigg(-\frac{\partial^2}{\partial \mathbf r_i^2} +
 \mathbf r_i^2 \bigg) \nonumber \\
&+ \lambda \sum_{i < j}^N \frac{1}{|\mathbf{r}_i - \mathbf{r}_j|}
 \Bigg] \mathrm \Psi(\mathbf r, t) \; ,
\end{align}
where \( l_0= \left( \hbar / \left( m\Omega\right)\right)^{-1/2} \) 
is the well-known oscillator length and 
\begin{equation}
 \lambda = \frac{mal_0}{\hbar^2}
\end{equation}
is the so-called coupling parameter. 
Due to the rescaling, there will only be dimensionless quantities
throughout this work. For example, lengths, times and energies 
are given in units of \(l_0\), \(\Omega^{-1}\) and \( \hbar \Omega\),
respectively.
The meaning of \(\lambda\) can be interpretated as follows.
Defining the scale of the potential energy as
\begin{equation}
 E_0 = \frac{1}{2}m\Omega^2 l_0^2
\end{equation}
and the mean interaction energy as
\begin{equation}
 E_\mathrm{C} = \frac{a}{2l_0} \;,
\end{equation}
one finds
\begin{equation}
 \lambda = \frac{E_\mathrm{C}}{E_0} \,.
\end{equation}
Hence, \( \lambda \) can be understood as the ratio of the interaction 
energy and the confinement energy. The influence of the value of 
\(\lambda\) is described later in Sec. \ref{sec:lambda}. The actual 
excitation of the breathing mode is realized by a fast switch of 
the trapping potential. For a short period of time \(t_\mathrm{off} \) 
the single-particle potential \( v \)  is completely turned off.
As a consequence, the time-dependent Hamiltonian takes the form
\begin{align}
\label{eq:tdhamiltonian}
\hat H(t) = &\sum _{i=1}^N \hat h_i + 
\left[ \theta(t_0-t) + \theta(t - t_\mathrm{off}) \right] v(\mathbf r_i) \nonumber  \\
&+ \sum_{i < j}^N w(|\mathbf{r}_i - \mathbf{r}_j|) \;.
\end{align} 
During the off-time the particles are driven out of their initial state. When the potential is restored, the time-dependent expectation value of some quantities start to oscillate. In particular for the breathing mode, the oscillation of the single-particle potential energy is dominated by a beating of two harmonic oscillations. The corresponding frequencies will be called \( \omega_r \) and \( \omega_R \) from now on. Both a typical time series of the potential energy and the excitation process are demonstrated in Fig. \ref{fig:zeitserie}. In the next subsections we want to point out some algebraically accessible properties of the frequencies and their relations to the coupling parameter \(\lambda\).

\subsection{Separation of center-of-mass motion}
As it has already been shown\cite{bauch}, the system always possesses a universal breathing mode whose frequency has the value \(\omega_R=2\). In the following, we will derive this result and show that this value is independent of the coupling strength, the system dimensionality and the particle number. 

The basic idea is the introduction of center-of-mass and relative coordinates for the solution of Eq. (\ref{eq:schroedrescaled}) with a product ansatz. The center-of-mass coordinate is given by
\begin{equation}
 \mathbf{R} \equiv \frac{1}{N} \sum_{i=1}^N \mathbf{r}_i\;,
\end{equation}
and the set of relative coordinates is given by
\begin{equation}
\mathbf{x} \equiv \left( \mathbf{x}_{1},\; \mathbf{x}_{2},\dots, \mathbf{x}_{N-1} \right)\;, 
\end{equation}
with the definition 
\( \mathbf{x}_{i} \equiv  \mathbf{r}_{i,i+1} \equiv \mathbf{r}_{i+1} -\mathbf{r}_{i} \). 
Thus, \( \mathrm{\Psi}(\mathbf{r},t) \) in
 Eq. (\ref{eq:schroedrescaled}) is replaced 
 by \(\mathrm{\Psi}(\mathbf{R},\mathbf{x} ,t)\). 
 Now the transformation is shown for each term in Eq.  (\ref{eq:schroedrescaled}), starting with
 \begin{align}
-\frac{1}{2}\sum_{i=1}^N \frac{\partial^2}{\partial \mathbf{r}_i^2} 
&=-\frac{1}{2}\left( \sum_{i=1}^N  \frac{\partial^2}{\partial \mathbf{R}^2} \frac{1}{N^2} +2\sum_{i=1}^{N-1} \frac{\partial^2}{\partial \mathbf{x}_{i}^2} \right) \nonumber \\
&=-\frac{1}{2N} \frac{\partial^2}{\partial \mathbf{R}^2} - \sum_{i=1}^{N-1} \frac{\partial^2}{\partial \mathbf{x}_{i}^2} \;.
\end{align}
For the second term we obtain 
\begin{align}
\label{eq:potenergie_falle}
\frac{1}{2}\sum_{i=1}^N \mathbf{r}_i^2 = \frac{1}{2}N\mathbf{R}^2 + \frac{1}{4N} \sum_{i=1}^N\sum_{k=1}^N \mathbf{r}_{ik}^2 \;,
\end{align}
where \( \mathbf{r}_{ik}^2  \) still has to be expressed in relative coordinates \( \mathbf{x} \). 
This can be be done as follows:
\begin{align}
\mathbf{r}_{ik}= 
\begin{cases}
  \displaystyle \sum_{l=i}^{k-1} \mathbf{x}_{l}    & \text{for } i < k \\
  \displaystyle \sum_{l=k}^{i-1} \mathbf{x}_{l}   & \text{for } i > k \\
  0    & \text{for } i = k \;.
 \end{cases}
\end{align}
Finally, the third term takes the form
\begin{align}
 \lambda \sum_{i<j}^N \frac{1}{|\mathbf{r}_i -\mathbf{r}_j|}
 = \lambda \sum_{i<j}^N \frac{1}{\left| \sum_{l=i}^{j-1} \mathbf{x}_{l}\right|}\;.
\end{align}
As a result the Hamiltonian can be split in two independent contributions:
\begin{align}
\hat{H}_\mathbf{R} = -\frac{1}{2N} \frac{\partial^2}{\partial \mathbf{R}^2} + \frac{1}{2}N\mathbf{R}^2
\end{align}
and
\begin{align}
\hat{H}_{\mathbf{x}} = &-\sum_{i=1}^{N-1} \frac{\partial^2}{\partial \mathbf{x}_{i}^2} + \frac{1}{4N} \sum_{i=1}^N\sum_{k=1}^N \mathbf{r}_{ik}^2 \nonumber \\
&+ \lambda \sum_{i<j}^N \frac{1}{\left| \sum_{l=i}^{j-1} \mathbf{x}_{l}\right|} \;.
\end{align}
Hence, the TDSE takes the form
\begin{align}
 \label{eq:tdse_trafo}
 \mathrm{i}\frac{\partial}{\partial t} \mathrm{\Psi}(\mathbf{R},\mathbf{x},t) = \left( \hat{H}_\mathbf{R} + \hat{H}_{\mathbf{x}} \right) \mathrm{\Psi}(\mathbf{R},\mathbf{x},t) \;.
\end{align}
The product ansatz
\begin{equation}
\mathrm{\Psi}(\mathbf{R},\mathbf{x},t) = \phi(\mathbf{R},t) \, \varphi(\mathbf{x},t) 
\end{equation}
yields the independent problems
\begin{equation}
\label{eq:tdse_sp}
\mathrm{i}\frac{\partial}{\partial t}\phi(\mathbf{R},t)= \hat{H}_\mathbf{R} \phi(\mathbf{R},t)
\end{equation}
and
\begin{equation}
\mathrm{i}\frac{\partial}{\partial t} \varphi(\mathbf{x},t) = \hat{H}_{\mathbf{x}} \varphi(\mathbf{x},t) \;.
\end{equation}
The center-of-mass problem can be transformed to the standard oscillator form
\begin{equation}
 \label{eq:tdse_sp_scaled}
\mathrm{i}\frac{\partial}{\partial t} \phi(\tilde{\mathbf{R}},t)=\left(-\frac{1}{2} \frac{\partial^2}{\partial \tilde{\mathbf{R}}^2} + \frac{1}{2}\tilde{\mathbf{R}}^2 \right) \phi(\tilde{\mathbf{R}},t)\;,
\end{equation}
where the rescaling \( \tilde{\mathbf{R}} = \sqrt{N} \mathbf{R} \) has been used.
Now consider an initial state, which can be expressed by
\begin{equation}
\label{eq:tdse_anfangsbed}
\phi(\tilde{\mathbf{R}},t=0)=\sum_n c_n \phi_n(\tilde{\mathbf{R}}) \;,
\end{equation}
where \( \phi_n \) is a solution of 
\begin{equation}
 \left( -\frac{1}{2} \frac{\partial^2}{\partial \tilde{\mathbf{R}}^2} + \frac{1}{2}\tilde{\mathbf{R}}^2 \right) \phi_n(\tilde{\mathbf{R}}) = E_n \phi_n(\tilde{\mathbf{R}}) \;.
\end{equation}
The associated energy eigenvalues are well-known: \( E_n = n+ d/2 \) for all \( n \in \{ 0,1,2,\dots \} \), and the time evolution of the state in Eq.  (\ref{eq:tdse_anfangsbed}) is given by 
\begin{equation}
\phi(\tilde{\mathbf{R}},t)=\sum_n c_n \phi_n(\tilde{\mathbf{R}}) \mathrm{e}^{-\mathrm{i}E_nt}\;.
\end{equation} 
The breathing mode manifests itself in the dynamics of the quantity \( \mathbf{r}^2 = \sum_{i=1}^N \mathbf{r}_i^2  \). Using center-of-mass and relative coordinates, this quantity can be expressed according to Eq. (\ref{eq:potenergie_falle}). For the first term in Eq. (\ref{eq:potenergie_falle}) a breathing frequency can be obtained by determining the expectation value 
\begin{equation}
\label{eq:exvalue}
\langle \mathbf{R}^2 \rangle(t) = N^{-1/2}\langle \tilde{\mathbf{R}}^2 \rangle(t) \;.
\end{equation}
The result for this expression is

\begin{align}
 \label{eq:ewbmqm}
 \langle \tilde{\mathbf{R}}^2 \rangle(t) 
&= \int \mathrm{d}\tilde{\mathbf{R}} \,\mathrm{d}\mathbf{x} \, \mathrm{\Psi}^*(\tilde{\mathbf{R}},\mathbf{x},t) \tilde{\mathbf{R}}^2  \mathrm{\Psi}(\tilde{\mathbf{R}},\mathbf{x},t) \nonumber \\
&= \sum_{i,j}c_i^*c_j\mathrm{e}^{-\mathrm{i}(j-i)t} ( \tilde{\mathbf{R}}^2)_{ij} \;, 
\end{align}
with 
\begin{align}
( \tilde{\mathbf{R}}^2)_{ij} = \int \mathrm{d}\tilde{\mathbf{R}} \, \phi_i^* (\tilde{\mathbf{R}}) \tilde{\mathbf{R}}^2  \phi_j(\tilde{\mathbf{R}}) \;.
\end{align}
It can be shown with a reduction to the matrix elements of a 
one-dimensional harmonic oscillator that only the cases 
\( i=j\) and \( i=j\pm2\) contribute to the last summation in Eq. (\ref{eq:ewbmqm}) 
(the case \( i=j\) does not correspond to an oscillation). 
The only frequency appearing in the oscillation is thus 
given by \( \omega_{{R}}=2 \). As the coupling parameter \(\lambda\)
does not appear in the center-of-mass problem, the center-of-mass mode with frequency \( \omega_{{R}}=2 \) is present for all couplings. In summary, we have shown that the frequency \( \omega_R \) is universal, but its amplitude tends to vanish for large particle numbers since it is proportional to \( N^{-1/2} \).

\subsection{Influence of coupling parameter and limiting cases}
\label{sec:lambda}
As we have seen in the last subsection, the introduction of relative and 
center-of-mass coordinates has led to a separation ansatz. 
The center-of-mass Hamiltonian \(\hat H_\mathbf{R} \) yields a breathing 
frequency \( \omega_R = 2\). It has already been mentioned that the 
breathing mode also exhibits another frequency \(\omega_r \). The properties 
of this frequency are an interesting subject of numerical analysis. Only in two
limiting cases the values of \(\omega_r\) are known from algebraic calculations. 
In the pure quantum limit, \( \lambda = 0\), the particles are completely uncorrelated.
As the interaction term in the Hamiltonian is cancelled out, a degenerate 
frequency \(\omega_r=\omega_R=2\) occurs. On the contrary, in the classical 
limit \( \lambda \to \infty\) the frequency \( \omega_r\) has 
the value \(\sqrt{3} \) \cite{peeters, dubin, henning}. In both cases the 
frequency does not depend on the particle number or the dimensionality of 
the system. For arbitrary coupling parameters the values of \( \omega_r \) 
are expected to be in the interval \(\left[\sqrt 3, 2 \right] \). To clarify 
the influnce of the system parameters such as the coupling parameter and the 
particle number on \( \omega_r\) is one of the main goals of our investigation.

\section{Simulation Methods}
Whereas the solution of the time-dependent Schr\"odinger equation could only be performed for 2 particles, the frequencies for up to 8 particles are still accessible through exact Configuration Interaction calculations.
These results are used to support the accuracy of time-dependent Hartree-Fock calculations, which have been performed for even higher particle numbers (up to 20).

Due to the high computational effort, we restrict ourselves to the solution of a 1-dimensional system. Nevertheless such a system can be regarded as a basic theoretical model which requires a deepened understanding. 
In order to suppress spin effects, only spin-polarized systems are investigated. Before the breathing mode is excited, the system is in the energetically lowest anti-symmetric state.

In the following, we give a brief discussion of the employed methods and present their numerical results. It shall already be mentioned here that in order to avoid divergencies in the interaction potential all methods use a regularized Coulomb potential 
\( \lambda / | \mathbf r_i - \mathbf r_j + \kappa ^2 | \), where \( \kappa \) is a small finite cut-off parameter.

\subsection{Time-dependent Schr\"odinger equation}
In our previous work \cite{bauch} we determined \(\omega_r(\lambda)\) for \(N=2 \) in the whole range \( \lambda = 0 \ldots \infty\). These values are the basis for the comparison with other methods. Our TDSE results were obtained by solving the time-dependent Schr\"odinger equation with two different methods. On the one hand, a standard grid-based Crank-Nicolson scheme was used, and on the other hand the wave function was expanded into a set of basis functions (oscillator eigenfunctions). The results confirm the values of the breathing frequencies in the limiting cases (\( \lambda = 0 \) and \( \lambda = \infty \)) and yield a continuous function \( \omega_r(\lambda) \) for all other couplings in between.

\subsection{Configuration Interaction}
Configuration Interaction (CI) is another method for obtaining numerically exact results. The basic idea of CI is to expand the wave function in a complete set of Slater determinants, which in turn are constructed with a complete single-particle basis. 
It has already been stated in Eq. (\ref{eq:tdhamiltonian}) that the excitation of the breathing mode is realized by a fast Heaviside-type switch-off of the single-particle confinement potential. Assuming the excitation to be infinitely short, the Hamiltonian is given by 
\begin{equation}
 \hat{H}(t) =	\hat{H}_0 + \eta \hat{H}_1 \delta(t) \;,
\end{equation}
where the time-independent part \(\hat{H}_0\) is that of Eq.  (\ref{eq:tidhamiltonian}). If \( \hat{H}_0 \) is diagonalized by the eigenfunctions \( | \Psi_n \rangle\) with eigenvalues \( E_n \), the application of perturbation theory yields that the expectation value of an arbitrary observable can be calculated by 
\begin{equation}
 \langle \hat{A} \rangle (t) = \sum_{i,j}  c^*_i c_j \: e^{\mathrm{i}(E_i - E_j) \,t } \: \langle \Psi_i | \hat{A} | \Psi_j \rangle \;,
\end{equation}
where \( c_{i,j} \) are time-independent coefficients. As a consequence of this relation, the oscillation of the expectation value is restricted to frequencies \( \omega_{ij} \equiv | E_i - E_j | \). 
Instead of time-propagating the solution of the Schr\"odinger equation, we can use this result and extract the breathing frequencies from the eigenvalues of \( \hat{H}_0 \) with relatively little computational effort. 
However, this method is only applicable for small particle numbers because of the strongly increasing size of the required basis sets for higher particle numbers. 
All results were produced with a basis of oscillator functions. Especially for weak couplings this basis set is well adjusted to the physical problem and the number of basis functions can be kept low. 
Just like the TDSE results the CI results can be used as a benchmark for the accuracy of the time-dependent Hartree-Fock results.


\subsection{Time-dependent Hartree-Fock}
For larger particle numbers (\(N \geq 7\)) approximation methods have to be employed, because the basis size for reasonable CI results dramatically increases. In the following, the interaction of the system is approximated on the mean-field level. The method of choice is the use of Nonequilibrium Green's functions $G^<(1,2)$, where $1\equiv(\mathbf{r}_1,t_1)$.  Employing the Hartree-Fock (HF) approximation, the Green's functions obey the Keldysh/Kadanoff-Baym equations \cite{bonitzintro, kadanoffbaym}
\begin{align}
\label{MS1}
\left(i \partial_{t_1}-h^\mathrm{HF}_1 (1)\right)G^<(1,2) &= 0 \;,\\
\label{MS2}
h^\mathrm{HF}_1(1) &= h^0(1)+h^\mathrm{HF}[G^<](1)\;,
\end{align}
where $h^0(1)$ is the single-particle part of the Hamiltonian (\ref{eq:tidhamiltonian}).
 In such a reduced quantum statistical description of the system all one-particle 
 quantities (and some \(N\)-particle quantities, e.g. the total energy) can be derived from the one-particle density matrix~\cite{bonitzteub} 
\begin{equation}
 F^1(\mathbf{r}_1,\mathbf{r}_2,t) = - \mathrm{i} \lim_{t' \to t^+ } G^<(\mathbf{r}_1,t,\mathbf{r}_2,t') \;,
\end{equation}
where the limit is taken from above. The advantage of the quantum statistical description is the fact that the particle number 
is only a parameter and does not determine the size of a whole system of equations as in an equivalent wave-function-based method.
The Green's function is expanded in FE-DVR (finite-element discrete variable representation) basis functions
 \cite{fedvr1, fedvr2} with up to $\sim$400 basis functions. The numerical procedure is to calculate the ground 
 state (with inverse temperature \(\beta\to\infty \)), switch off and on the confinement and continue propagating 
 the Green's function in time. In each step the expectation values $\langle E_\mathrm{pot} \rangle$ are calculated and saved. 
 Finally, one can analyze the spectrum of the time series and extract the breathing frequency out of it.
Compared to methods based on perturbation theory, which only involve the ground state energies, this method is quite time consuming. 
Since the computation of just a single frequency can last more than one day on one CPU, the paramaters have to be chosen carefully in order to guarantee converged results 
while keeping the durations of the computations acceptable.


\section{Results}

As mentioned before, we want to concentrate on presenting the results of our time-dependent Hartree-Fock (TDHF) calculations and show the dependency of \( \omega_r \) on the particle number. In this work we only investigate the cases \(\lambda~=~0.1\), \(0.3\) and \(1\). Larger values of \( \lambda \) would require to go beyond Hartree-Fock, which demands a very high computational effort. Before we show our results, we start by explaining some important aspects concerning the spectral determination of the breathing frequencies. 

\begin{figure}[b]
 \begin{center}
  \includegraphics[scale=1]{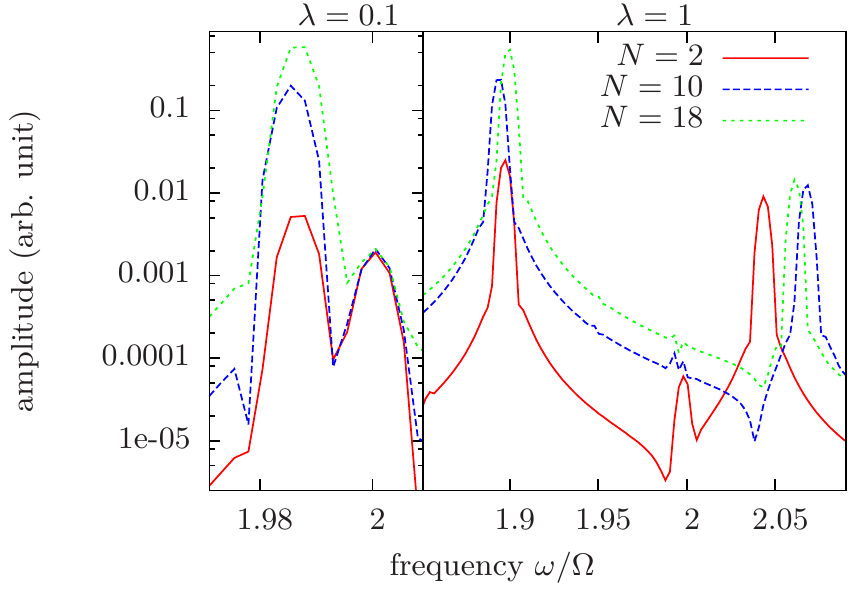} 
  \end{center}
  \caption{Comparison of the spectra for 2, 10 and 18 particles at different coupling strengths \( \lambda \). The left peaks represent the relative mode with frequency \( \omega_r \). Peaks with slightly higher frequencies than the center-of-mass frequency \( \omega_R = 2 \) can be found in case \( \lambda=1 \). For \( \lambda=0.1 \) these peaks and the center-of-mass peaks might overlap.}
  \label{fig:spectra}
\end{figure}

\begin{figure}[t]
 \begin{center}
  \includegraphics[scale=1]{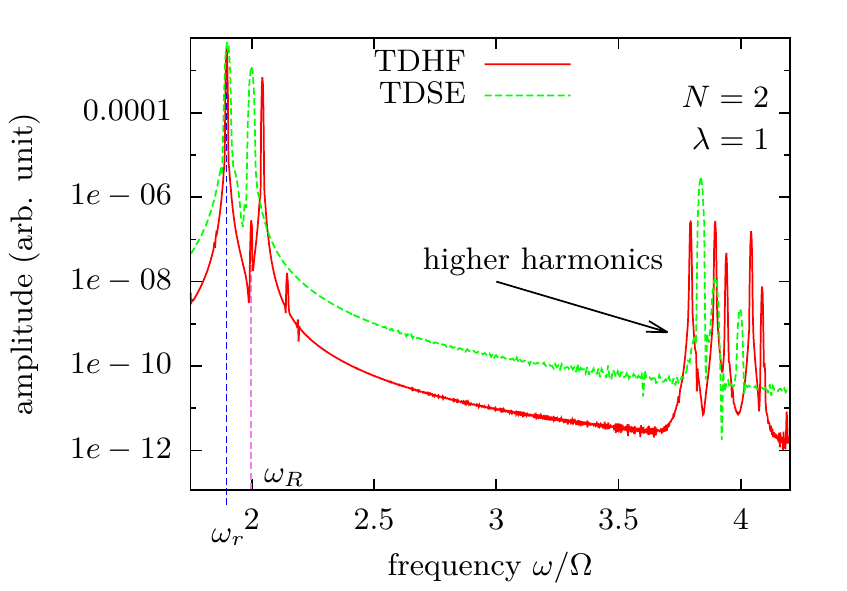} 
  \end{center}
  \caption{Spectrum corresponding to time series in Fig. \ref{fig:zeitserie} (TDHF) in comparison with the spectrum of the solution of the Schr\"odinger equation (TDSE) for the same configuration. The two breathing frequencies \( \omega_r \) and \( \omega_R \) can be identified in both spectra. However, the TDHF spectrum exhibits numerous additional peaks. One of those (unphysical) peaks is near \(\omega = 2\) and has a high amplitude. Aside from that both spectra show higher harmonics, which, however, do not agree with each other.}
  \label{fig:spekges}
\end{figure}
\subsection{Spectral Analysis}
The spectra of all TDHF results were calculated from time series which have the length of at least \( t_\mathrm{prop} =2000\) 
(in units \(\Omega^{-1}\)). Since the resolution in the frequency space is limited by the size of \( t_\mathrm{prop} \),
 we applied spline interpolations to the spectra in order to achieve a higher accuracy for the frequency values.
 Besides, each spectrum was calculated with a Blackman window in order to uncover peaks with small spectral weights. 
 In Fig. \ref{fig:spectra} the interesting part of the spectra around the breathing 
 frequencies is shown for different couplings and particle numbers. The peaks corresponding to \( \omega_r \) can be 
 clearly identified. However, the identification of the center-of-mass peaks corresponding to 
 \( \omega_R = 2 \) is problematic. Only in the case \( \lambda = 1 \) it is evidently possible 
 to find peaks at frequency \( \omega = 2 \), although they have a very small spectral weight and tend to vanish with
increasing particle number. Surprisingly, peaks with slightly higher frequencies than \( \omega = 2 \) and 
a rather strong dependency on the particle number are observed. As their spectral weight is some orders of magnitude larger than that of the center-of-mass peaks,
one has to be careful not to confuse those peaks with the center-of-mass peaks. This is important when regarding the spectra in cases 
\( \lambda = 0.1 \) and \( \lambda = 0.3 \), because for these couplings the small peaks 
at \( \omega = 2 \) vanish. It is conceivable that these peaks and those slightly higher than \( \omega = 2 \) tend to overlap.

The question remains why the spectra show such strong additional peaks. In Fig. \ref{fig:spekges} the spectra 
of both TDSE and TDHF calculations are shown. Both spectra were produced with the same parameters 
(\( \lambda=1, N=2, \kappa^2 = 0.1\)). As it can be seen, the TDHF spectrum contains several additional peaks. 
It turns out that the appearance of (unphysical) frequencies is a typical effect of the TDHF approximation. 
Consequently, it can be a challenging task to distinguish between real physical frequencies and artefacts of 
the approximation method. However, in both spectra the peaks of higher harmonics occur (Only the first higher 
harmonics can be seen in the figure, but the full data even contain higher harmonics). They are caused by the finite duration 
of the initial excitation. Unfortunately, the values of the higher harmonics are not in a good agreement with each other.
Taking into account that the spectral weight of the higher harmonics is quite small, it is presumable that the accuracy 
of the calculations is not high enough to properly represent the higher harmonics. In the following, we want to focus on
the breathing frequency \( \omega_r \) and neglect all higher order spectral features.

\begin{figure}[t]
  \begin{center}
  \includegraphics[scale=1]{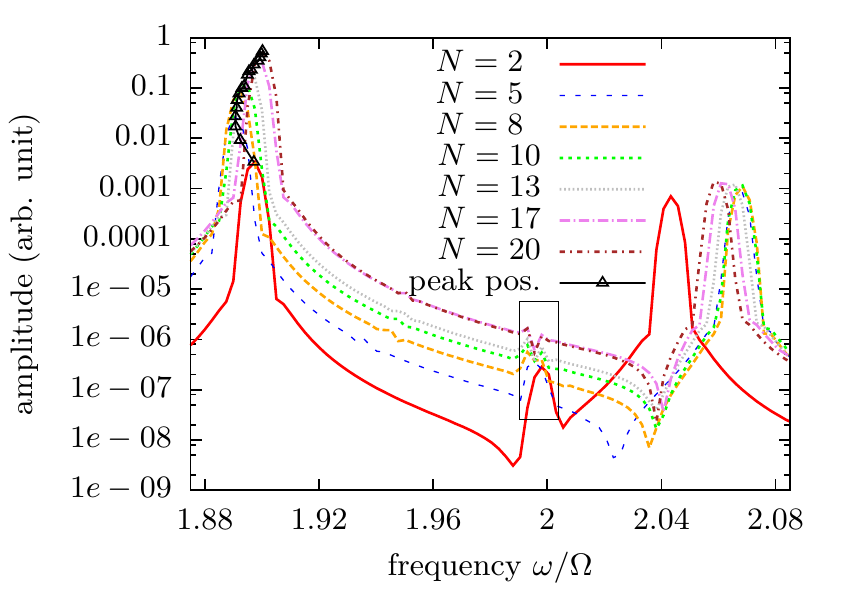} 
  \end{center}
    \caption{Part of the spectrum around the breathing frequencies for different particle numbers at coupling strength \( \lambda = 1 \). While the relative mode frequency \( \omega_r \) slightly changes with $N$ (black line), the center-of-mass mode  frequency \( \omega_R\) (black rectangle) remains constant (apart from numerical errors). Note the strongly \(N\)-dependent additional peaks on the right, which must not be confused with the \( \omega_R\) peaks.}
    \label{fig:spekN}
\end{figure}

\begin{figure}[t]
  \begin{center}
  \includegraphics[scale=1]{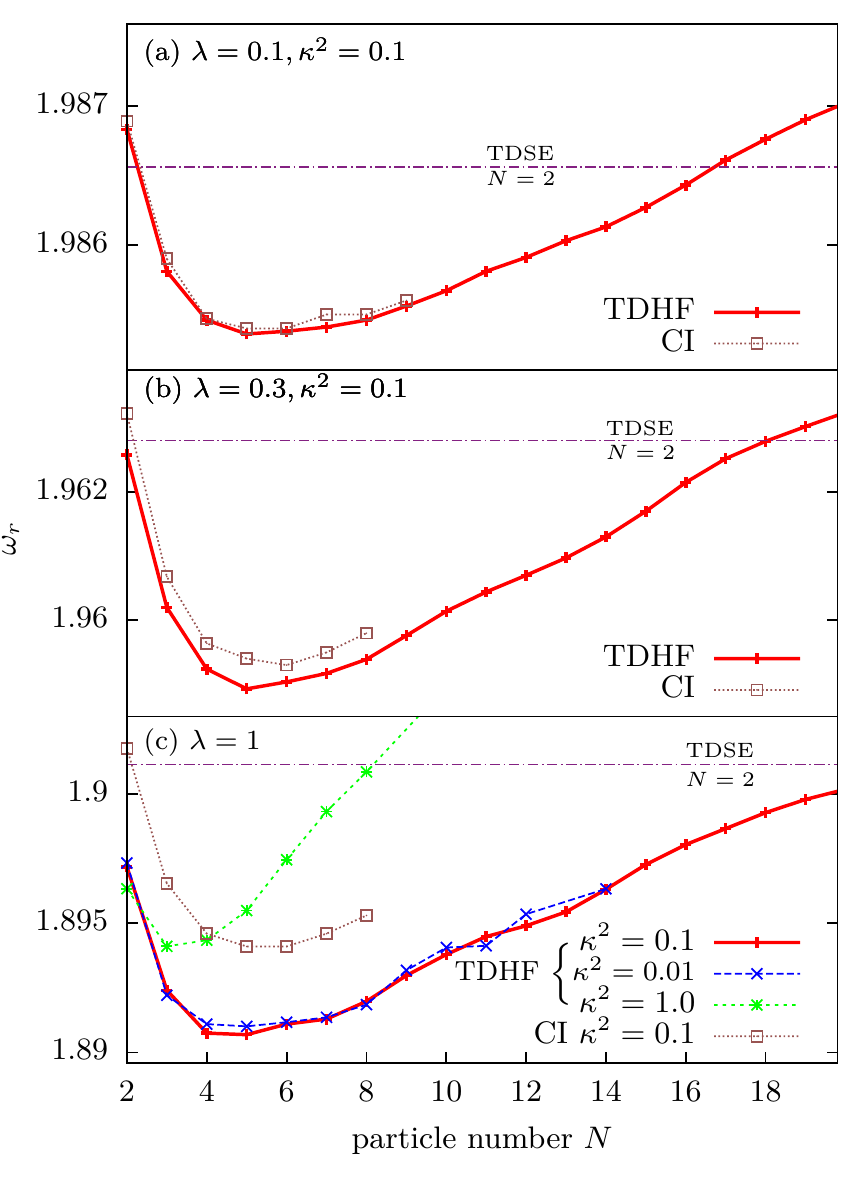} 
  \end{center}
    \caption{Breathing frequency $\omega_{r}$ vs. particle number $N$ for coupling parameters \(\lambda~=~0.1\), \(0.3\) and \(1\) with cut-off parameter \(\kappa^2=0.1\). For \( \lambda = 1 \) the frequencies obtained for different \(\kappa^2\) are also plotted. For comparison, the TDSE frequencies for two particles and the CI results are plotted as well.
For \(\kappa^2=0.1\), the \(N\)-dependencies are qualitatively the same for all \( \lambda\).}
    \label{fig:omega_lambda}
\end{figure}
\subsection{$N$-dependency of the breathing frequency}
The TDHF calculations enabled us to obtain the breathing frequencies for up to 20 particles. Apart from the problem of clearly identifying the center-of-mass mode, 
we found a typical behaviour of the breathing frequency \( \omega_r \), which is qualitatively the same for all couplings.  
The change of the spectra with increasing particle number is illustrated in Fig. \ref{fig:spekN}. It is obvious that the breathing frequency \( \omega_r \)
decreases to a minimum (for 5 particles, see also Fig. \ref{fig:omega_lambda}) and afterwards monotonically increases. 
The figure also shows the center-of-mass peaks, which, as expected from Eq.  (\ref{eq:exvalue}), tend to vanish for high particle numbers.
Furthermore, the additional peaks near \( \omega = 2 \) exhibit a rather strong \( N\)-dependency, but appear to be converged.

After analyzing the spectra for the couplings \(\lambda~=~0.1\), \(0.3\) and \(1\), it turns out that the \( N\)-dependency of \( \omega_r\) 
is qualitatively equal for all \( \lambda\).
It is typical that the frequencies attain a minimum for 5 particles, before they start steadily growing.
For a complete overview the explicit values of the breathing frequencies are summarized in the graphs of Fig. \ref{fig:omega_lambda}. For comparison, 
we also show the TDSE values for \(N=2\) and the CI values for up to 
8 particles as well as the breathing frequencies obtained with other cut-off parameters. 
As expected, the TDHF and the CI values are almost the same (see Fig.
 \ref{fig:omega_lambda}(a)) for small \(\lambda\). 
 With increasing \(\lambda\) a constant shift between both results arises. 
 The CI results confirm that the breathing frequency has a minimum, which, however, occurs for 6 particles instead of 5 particles in the case of TDHF.
Moreover, the variation of \( \kappa \) indicates that the results are converged in the 
regime \( \kappa^2 = 0.01 \ldots 0.1 \), because the results just marginally differ from each other.

\begin{figure}[t]
  \begin{center}
  \includegraphics[scale=1]{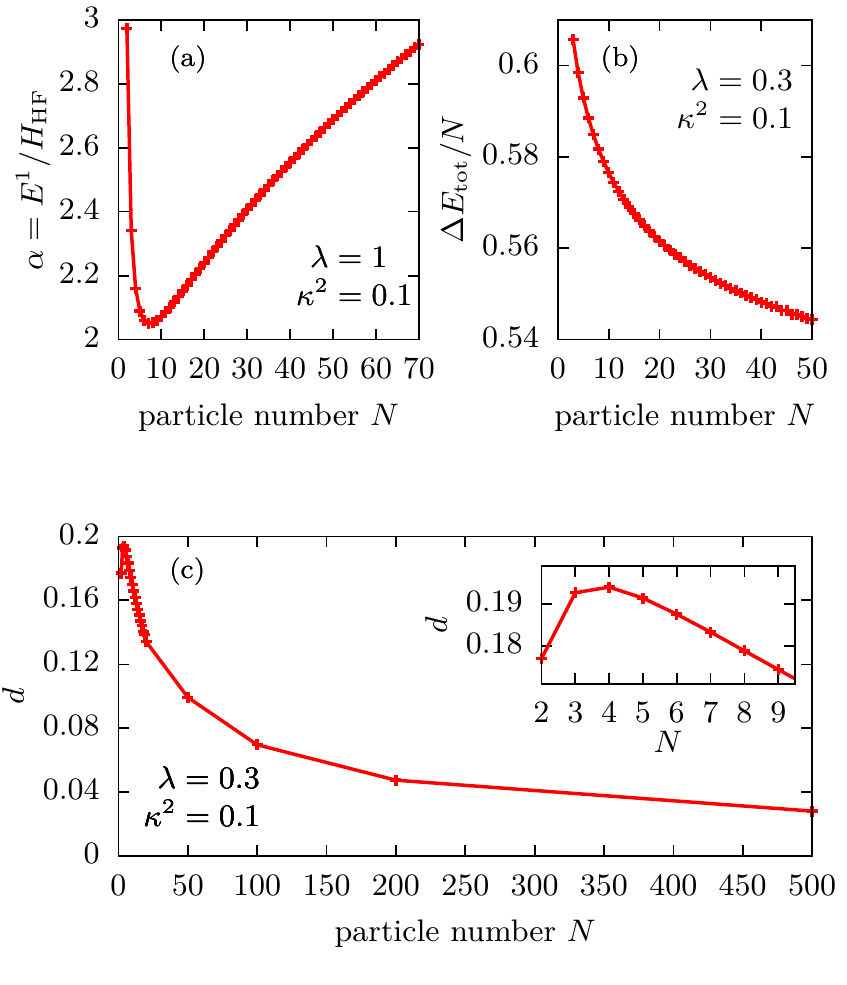} 
  \end{center}
  \caption{\(N\)-dependent ground state properties of the interacting system. The inverse coupling parameter \(\alpha\) has a minimum for 7 particles and afterwards increases with $N$ (a).
  The difference of the addition energy continuously decreases and possibly reaches the ideal value 0.5 for large particle numbers (b). The pointwise deviation of
  the densities $d$ has a maximum for 4 particles and tends to vanish for large particle numbers (c).}
  \label{fig:indicators}
\end{figure}
For 20 particles the frequency \( \omega_r \) has not yet reached an asymptotic value. There are, however, several indicators that for \( N \to \infty\) 
the frequency \( \omega_r \) converges to the value \( \omega = 2 \) characteristic for the ideal quantum limit. In the following, we just consider the ground state. 
This assumption is justified by the fact that the breathing mode is computed in linear response, which should be derivable from the ground state properties.

Firstly, if the system converges to a pure quantum state, the interaction energy must become a negligible quantity. 
Assuming that the Hartree-Fock part of the total energy represents the interaction energy, 
it is possible to calculate the ratio of the one-particle energy \( E^1\) to the interaction energy \(E_\mathrm{HF}\):
\begin{align}
 \alpha = \frac{\langle E^1 \rangle}{\langle E_\mathrm{HF}\rangle} = \frac{\langle E_\mathrm{kin}\rangle+\langle E_\mathrm{pot}\rangle }{\langle E_\mathrm{HF}\rangle} \;.
\end{align}
This quantity can be interpretated as the inverse of an effective coupling parameter.
The values of \(\alpha \) for up to 70 particles are shown in Fig. \ref{fig:indicators}(a). As expected, \(\alpha \) 
increases with the particle number. Interestingly, \(\alpha \) has a minimum for 7 particles.

Secondly, one can easily show that for an ideal quantum system the total energy per particle increases by \(0.5\) each time a particle is added to the system, i.e.
\begin{align}
\frac{ \Delta E_\mathrm{tot}}{N} \equiv \frac{E^N_\mathrm{tot} - E^{N-1}_\mathrm{tot}}{N} = 0.5 \;.
\end{align}
Calculating this addition energy for the interacting system, we get a result which is presented in Fig. \ref{fig:indicators}(b). The values slowly converge against the ideal value \(0.5\), 
although still larger values of $N$ would be required to prove this limit.

\begin{figure}[tb]
  \begin{center}
  \includegraphics[scale=1]{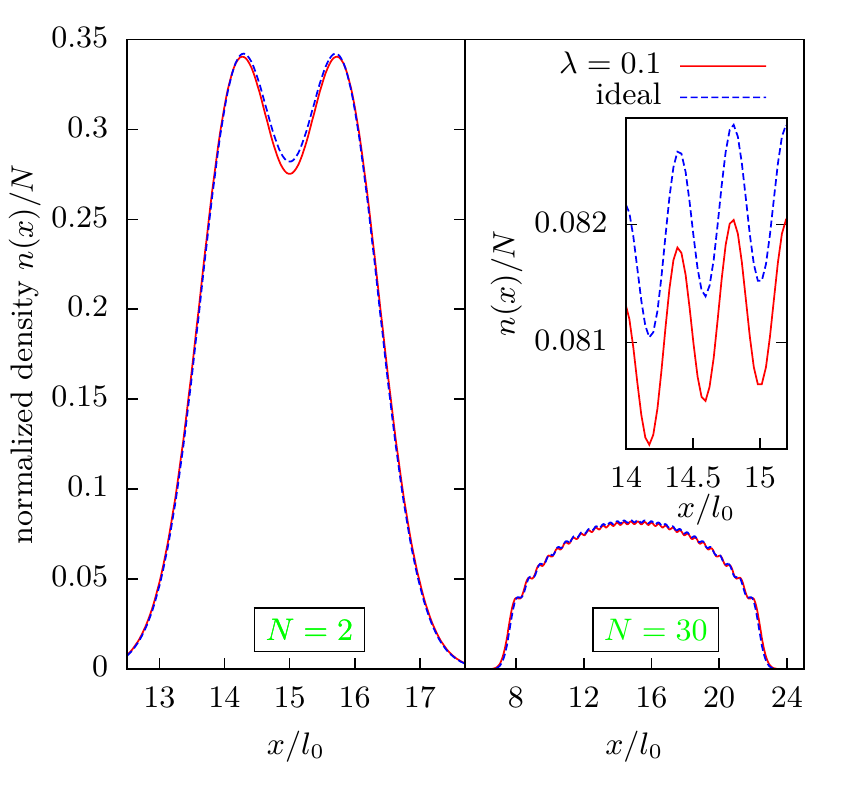} 
  \end{center}
  \caption{Comparison of the normalized densities of an ideal quantum system and an interacting system ($\lambda=0.1$) for 2 and 30 particles. The deviation from the ideal system vanishes for large particle numbers.}
  \label{fig:densities}
\end{figure}

A third indicator is the particle density \( n(\mathbf r) \) (or in the one-dimensional system \( n(x) \)). We want to show that 
for \(N \to \infty\) the density converges to the density of the ideal system. An appropriate quantity for that purpose is the pointwise
squared deviation of the normalized densities, i.e.
\begin{align}
 d \equiv \frac{1}{N}\sqrt{ \sum_{i} \left( n_\mathrm{ideal}(x_i)-n_\mathrm{interacting}(x_i)\right)^2  } \;.
\end{align}
The factor \(1/N \) prevents this quantity from diverging and gives it a relative character. (Recall the normalization
\( \int n(x) \mathrm{d}x = N\).) The values for up to 500 particles are shown in Fig. 
\ref{fig:indicators}(c). As expected, $d$ tends to vanish for large particle numbers and noticeably has a maximum for 4 particles. In order to demonstrate the change in the densities, the normalized densities of the ideal quantum system and the interacting system are plotted for 2 and 30 particles in Fig. \ref{fig:densities}.
Although all of the indicators are still not able to prove that for \(N \to \infty\) the system converges to an ideal quantum system,
they give an evident hint to this hypothesis.
An explanation for the transition to a pure quantum system might be that for large particle numbers the 
system reaches an extension which is dominated by the trap potential (\(\propto \mathbf{r}_i^2 \)). This is a consequence of the Pauli exclusion principle for fermions. Moreover, it is apparent that
the quantities \(\alpha \) and \(d\) have an extremum near the minimum of the breathing frequency. Yet it is not possible to clearly reconstruct
that mimimum from the ground state.

\subsection{Conclusions}
We showed that the breathing frequency \( \omega_r \) bears an \(N\)-dependency which is similar for the
 coupling strengths \(\lambda=~0.1\), \(0.3\) and \(1\). Presumably, the same behaviour can be observed for arbitrary 
 finite couplings. We found that the breathing frequencies decrease to a minimum 
 for 5 particles ($N=6$, for CI) and afterwards monotonically increase.
 
The origin of the minimum of $\omega_r$ is the competition between two
restoring forces: first, the Coulomb repulsion between
the particles which is strongest for $\lambda \to \infty$ (point charge)
and which favours $\omega_r = \sqrt{3}$. Second, the kinetic
energy of a quantum system that is maximal for $\lambda=0$. The minimum
appears where the ratio of the two, $1/\alpha$ has
its maximum. It may be expected that the occurence of this maximum is
related to particularly stable cluster configurations.

 Although we did not investigate more than 20 particles, we found some hints that for \(N\to\infty\) the breathing frequency 
 converges to the frequency \( \omega = 2 \) of an ideal quantum system. 
In contrast to the increase of $\omega_r$ with $N$ at finite $\lambda$,
for $\lambda \to \infty$, the frequency is independent
of $N$ and equals $\sqrt{3}$. Thus, presumably, there exists a critical
value of $\lambda$ where the crossover between the
two behaviours occurs.
It remains a subject of further research activities
 to confirm these hypotheses. Furthermore, it would be interesting to extend this analysis to higher dimensions,
 other interactions (e.g. dipole interaction) and other spin properties.

\bibliography{qbm}

\end{document}